\documentclass[twocolumn,prb,superscriptaddress]{revtex4}

\usepackage{graphicx}
\usepackage{dcolumn}
\usepackage{bm}
\usepackage{url}

\makeatletter
\newcommand\textsubscript[1]{\@textsubscript{\selectfont#1}}
\def\@textsubscript#1{{\m@th\ensuremath{_{\mbox{\fontsize\sf@size\z@#1}}}}}
\newcommand\textbothscript[2]{%
  \@textbothscript{\selectfont#1}{\selectfont#2}}
\def\@textbothscript#1#2{%
  {\m@th\ensuremath{%
    ^{\mbox{\fontsize\sf@size\z@#1}}%
    _{\mbox{\fontsize\sf@size\z@#2}}}}}
\def\@super{^}\def\@sub{_}

\begin{document}



\title{Hybrid Quantum Dot-2D Electron Gas Devices for Coherent Optoelectronics} 

\author{F. Dettwiler}
\affiliation{Department of Physics, University of Basel, CH-4056 Basel, Switzerland}

\author{P. Fallahi}
\affiliation{Institute of Quantum Electronics, ETH-Z\"urich, CH-8093 Z\"urich, Switzerland}

\author{D. Scholz}
\affiliation{Institut f\"ur Experimentelle und Angewandte Physik,~Universit\"at~Regensburg,~93040~Regensburg,~Germany}

\author{E. Reiger}
\affiliation{Institut f\"ur Experimentelle und Angewandte Physik,~Universit\"at~Regensburg,~93040~Regensburg,~Germany}

\author{D. Schuh}
\affiliation{Institut f\"ur Experimentelle und Angewandte Physik,~Universit\"at~Regensburg,~93040~Regensburg,~Germany}

\author{A. Badolato}
\affiliation{Institute of Quantum Electronics, ETH-Z\"urich, CH-8093 Z\"urich, Switzerland}
\affiliation{Department of Physics and Astronomy, University of Rochester, Rochester, NY14627, USA}

\author{W. Wegscheider}
\affiliation{Institut f\"ur Experimentelle und Angewandte Physik,~Universit\"at~Regensburg,~93040~Regensburg,~Germany}
\affiliation{Solid State Physics Laboratory, ETH-Z\"urich, CH-8093 Z\"urich, Switzerland}

\author{D.~M.~Zumb\"uhl}
\email[]{dominik.zumbuhl@unibas.ch}
\affiliation{Department of Physics, University of Basel, CH-4056 Basel, Switzerland}

\date{\today}
\begin{abstract}
We present an inverted GaAs 2D electron gas with self-assembled InAs quantum dots in close proximity, with the goal of combining quantum  transport with quantum optics experiments. We have grown and characterized several wafers -- using transport, AFM and optics -- finding narrow-linewidth optical dots and high-mobility, single subband 2D gases. Despite being buried 500\,nm below the surface, the dots are clearly visible on AFM scans, allowing precise localization and paving the way towards a hybrid quantum system integrating optical dots with surface gate-defined nanostructures in the 2D gas.
\end{abstract}


\maketitle
Spin qubits in gate-defined GaAs quantum dots \cite{LossDiVincenzo} are currently among the most promising candidates for a quantum processor \cite{HansonRevModPhys}, benefiting from excellent in-situ tunability and flexibility of gate-defined nanostructures in a GaAs 2D electron gas. However, due to a lack of hole confinement, these qubits do not easily couple to photons\cite{Pioda:2010,Fujita:2013}. Self-assembled quantum dots (SAQDs) e.g. made from InAs, on the other hand, are among the best solid-state single-photon emitters known today, acting as a spin-photon interface \cite{yilmaz:2010,Gao:2012,DeGreve:2012} and further allowing ultrafast and coherent optical manipulation \cite{BerezovskyAwschalom, PressYamamoto}. Placing a SAQD in close proximity to a surface-gate defined double dot can create a hybrid quantum system, opening the door for coherently converting the stationary spin qubit to a photon\cite{engel:2006}, and vice-versa -- \emph{coherent optoelectronics} -- thus enjoying the advantages of both systems in a combination of quantum optics and quantum transport experiments.

\begin{figure}[t]
\includegraphics[width=1\columnwidth]{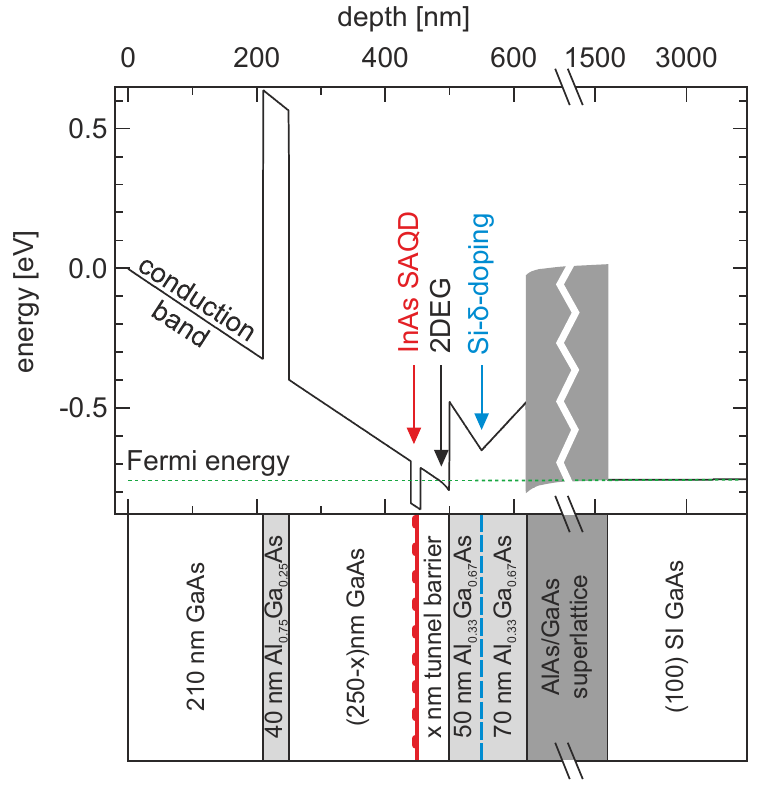}\vspace{-3mm}
\caption{{\bf Simulated conduction band energy (top) and wafer growth profile (bottom).} The tunnel barrier width $x$ between InAs dots and 2DEG is $15$,\,$30$,\,$45$ or $60$\,nm, allowing tunneling between 2DEG and dots. A blocking barrier halfway between surface and 2DEG suppresses leakage currents. The superlattice prevents formation of a mobile parallel conduction layer at the lower interface (at 620\,nm). The conduction band is calculated self-consistently\cite{Aquila}.\label{fig:conduction}}\vspace{-4mm}%
\end{figure}

InAs SAQDs in vicinity of a 2DEG (Fig.\,\ref{fig:conduction}) were shown to induce scattering\cite{Sakaki1995, Kim1998, Ribeiro1999, Kawazu2004, Kannan2007}, were studied with capacitance spectroscopy\cite{Russ2006, Lei2008} and were used for charge storage\cite{Finley1998}. Transport spectroscopy of a few or even single SAQDs in 2DEG structures were also done\cite{horiguchi1997, Vdovin2007, marquardt2008}. High-mobility 2DEGs in sufficiently close proximity to enable tunneling to narrow-linewidth optical dots have so far not been demonstrated. Designing an appropriate structure and optimizing its growth presents one important hurdle on the challenging path to coherent optoelectronics.

In this Letter, we present high quality inverted GaAs 2DEGs with a layer of narrow-linewidth optical InAs dots separated by a tunnel barrier of thickness 15\,nm\,$\leq\,x\,\leq$\,60\,nm. We characterize these structures with quantum transport measurements, atomic force microscopy (AFM) and optical spectroscopy. The mobility of the 2DEG depends on the tunnel barrier width, displaying higher mobilities for the thicker barriers. Further, the deformation field of the quantum dots is clearly visible on the surface with an AFM, even though the dots are buried roughly $500$\,nm below the surface. This allows localization of an individual quantum dot with 30\,nm accuracy, paving the way for hybrid systems integrating quantum dots in surface-gated nanostructures in the 2DEG.

The samples were grown on semi-insulating (100) GaAs substrates by molecular beam epitaxy (MBE), see Fig.\,\ref{fig:conduction} (lower panel) for the growth profile. The 2DEG is located $500$\,nm below the surface and is inverted, i.e. the Si $\delta$-doping layer is $50$\,nm below the 2DEG. An InAs wetting layer is grown above the 2DEG, separated by a tunnel barrier of width $15\,$nm\,$\leq x \leq 60$\,nm. During InAs growth the wafers were not rotated, resulting in a gradient of InAs thickness across the wafer. If the wetting layer exceeds $\sim1.6$ monolayer, Stranski-Krastanov InAs SAQDs -- henceforth referred to as \emph {dots} -- start to grow, with higher areal density for larger InAs thickness. Thus, a gradient of areal dot density develops across the wafer. The quantum dots are annealed in situ when partially capped with GaAs\cite{Hennessy:2007} to tune the ground state energy to $\sim$1.3\,eV. Successful dot formation, convenient emission wavelength and narrow linewidths are confirmed by $\mu$-PL gate-scans, which also reveal that the dots are neutral at zero gate-bias and start charging above $\gtrsim$0.2\,V.

We present the results of two different types of wafers: low-$T$ wafers, with InAs grown at $T\approx517\,^{\circ}$C, and high-$T$ wafers, with InAs grown at $T\approx534\,^{\circ}$C -- a slightly higher growth temperature aiming for better overgrowth and less defects. We note that for the InAs deposition at both low-$T$ and high-$T$, the temperature is significantly lowered for proper dot formation from the optimal GaAs growth temperature. Further, the low-$T$ wafers have an additional layer of as-grown dots of comparable density added on the surface. This provides a quick way to measure the dot density, e.g. with an AFM, without influencing the 2DEG because of its large depth. Only the buried dots close to the 2DEG are ultimately of interest here, allowing electron tunneling to the 2DEG and the desired optical activity, as needed for the outlined hybrid systems.

An inverted 2DEG structure was chosen because of two reasons. First, the strain field of the InAs wetting layer is expected to have a smaller effect on the 2DEG mobility when the dots are located above rather than below the 2DEG (dopants and dots should not be near each other on the same side of the 2DEG). Second, the alignment of the discrete energy levels of the quantum dots with the Fermi energy requires a relatively small external potential when the $\delta$-doping is below the 2DEG. This can be seen in the self-consistent numerical solution of the Schr\"odinger and Poisson equations\cite{Aquila} for our structure, delivering the conduction band energy as shown in Fig.\,\ref{fig:conduction} (upper panel) and showing that Fermi and lowest dot levels are indeed nearly aligned, consistent with $\mu$-PL gate-scans.

Additional difficulties arise from inverted heterostructures: the mobility is generally expected to be lower than in comparable normal (non-inverted) structures\cite{Morkos:1982,Kim:1990}, which have the Si atoms above the interface (in growth direction). This mobility degradation is predominantly due to migration of Si dopant atoms occurring mainly in the crystal growth direction\cite{pfeiffer:1991}. Nevertheless, when a sufficiently large undoped setback is incorporated, rather high quality single heterostructures\cite{pfeiffer:1991} and ultra-high mobility double heterointerface doped structures have been demonstrated\cite{Schlom:2013, Umansky:2013, Bockhorn:2013}. Here, we use a setback of 50\,nm, larger than the $\sim$30\,nm exponential Si-tail\cite{pfeiffer:1991}. Since the dots exclude another doping layer nearby as required for surface compensation, the interface is buried deep below the surface, at depth 500\,nm. Another undesirable effect is the formation of a parallel conduction layer at the lower AlGaAs/GaAs interface (see Fig.\,\ref{fig:conduction} at 620\,nm depth). This issue can be resolved with an AlAs/GaAs superlattice, extending from the lower interface for several hundred nanometers down into the wafer.

\begin{figure}[t]
\centering
\includegraphics[width=1\columnwidth]{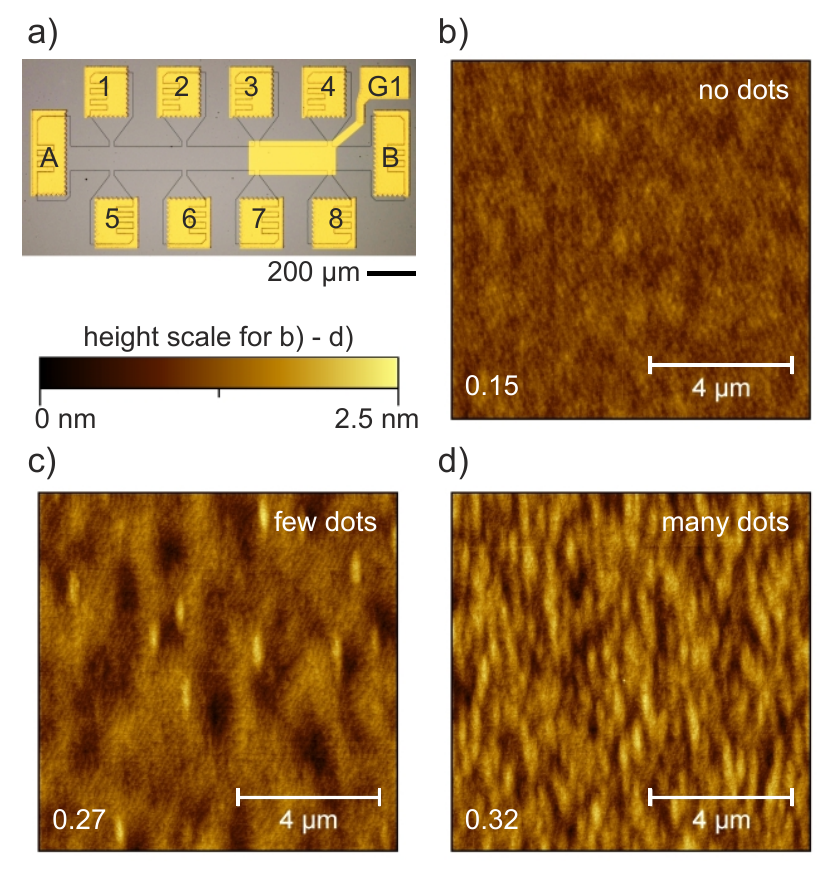}\vspace{-3mm}
\caption{\label{fig:AFM}{\bf Hall bar and buried-dot AFM measurements. (a)} Optical image of Hall bar sample.{\bf (b)-(d)} AFM images of the wafer surface with dots buried at a depth of $\sim$500\,nm (no dots directly on surface). {\bf(b)} Region without dots, {\bf(c)} with few dots, and {\bf(d)} with many (overlapping) dots. On the surface, the buried dots appear as cigar shaped hills of typical lateral dimension 800\,nm\,$\times$\,250\,nm oriented along the main crystal axes and a height of $\sim$1\,nm. Center of a hill can be determined to $\lesssim$30\,nm accuracy. RMS roughness values (lower left in (b)-(c)) also reflect the difference in dot density.}\vspace{-4mm}
\end{figure}

AFM imaging is the method of choice for accurately locating dots\cite{Hennessy:2007} with respect to other surface structures such as a marker grid. The deformation field of the InAs islands propagates upwards through the crystal, making it possible to image individual dots by AFM on the surface of the wafer, see Fig.\,\ref{fig:AFM}b)-d), despite a capping layer of about $500$\,nm. On the surface, the buried dots appear as cigar-shaped hills of much larger lateral size (compared to the actual size of the buried dot) and height of only $\sim$\,1\,nm. We can determine the dot position with an accuracy of $\lesssim$30\,nm from the AFM scans, assuming the center of the surface hill coincides with the buried dot, as is expected. Great care has to be taken to clean the surface after fabrication steps such as ebeam or optical lithography, making it challenging to locate these shallow hills. Previously, buried dots were localized in a similar manner but were situated only $\sim$70\,nm below the surface\cite{Hennessy:2007}. We note that for a very large number of dots, only a lower bound can be given due to overlapping hills.

\begin{figure}
\centering
\includegraphics[width=1\columnwidth]{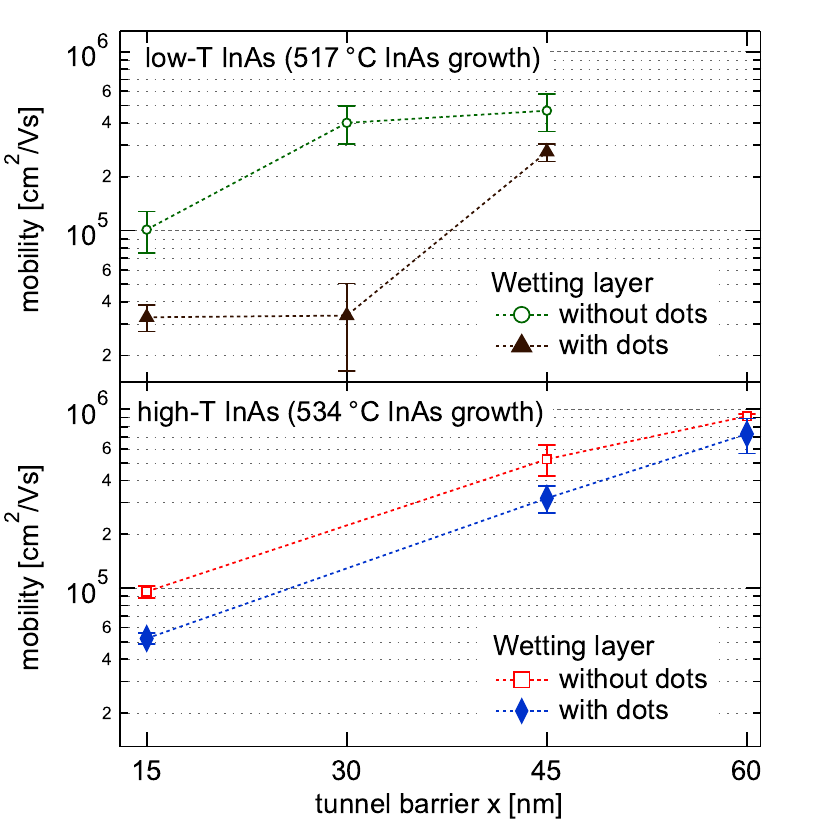}\vspace{-2mm}
\caption{{\bf Mobility as a function of tunnel barrier width} for low-$T$ and high-$T$ wafers (measured in the dark), shown over the same axis range in upper / lower panels for ease of comparison. For each data point, at least two different Hall bars were measured, and we average $\mu$ from gated and ungated regions (exhibiting no systematic difference); the statistical uncertainty gives the error bars. For each wafer, we compare high dot density devices (solid markers) to devices with only a thin InAs wetting layer without dot formation (empty markers). Low-$T$ 60\,nm and high-$T$ 30\,nm wafers were not grown. A small tunnel barrier width $x$ reduces the mobility significantly, even in absence of dots. For smaller $x$, the dots induce an additional suppression of $\mu$. The temperature during InAs growth affects the quality of the 2DEG only weakly. Dashed lines serve as a guide to the eye.\label{fig:mobility}} \vspace{-4mm}
\end{figure}

The 2DEG charge carrier density $n$ was extracted from the classical Hall slope and the mobility $\mu$ from the zero-field longitudinal resistivity measured on Hall bar samples, see Fig.\,\ref{fig:AFM} for layout. Measurements were done at $4.2$\,K with a standard 4-wire lock-in technique. The density $n\approx1.2\cdot10^{11}$\,cm$^{-2}$ is similar in all samples, as expected for a constant doping setback in all wafers. In Fig.\,\ref{fig:mobility}, the mobility is shown as a function of tunnel barrier width $x$, comparing  samples without dots and with large dot density, and further contrasting low-$T$ (upper panel) and high-$T$ (lower panel) InAs growth samples. As evident in Fig.\,\ref{fig:mobility}, a general drastic reduction of $\mu$ is observed for small tunnel barrier widths $x$.

We identify two scattering mechanisms (noting that the dots are charge neutral here): first, the presence of the InAs layer itself and second, crystal defects in the GaAs, enhanced by the temperature lowering required for the InAs growth. This was further explored with additional reference samples which were grown identically (including growth temperature profile) except that no InAs was deposited. Despite the lack of the lattice mismatched layer, the mobility of such wafers also exhibits a similar reduction as observed with InAs deposition in absence of dots, albeit slightly weaker. Thus, already the briefly lowered growth temperature is reducing the mobility and results in a more severe mobility reduction at smaller $x$. When the temperature reduction is omitted altogether, a high mobility is seen ($1.2\times10^6\,\mathrm{cm^2/(Vs)}$), slightly above the 60\,nm barrier samples, as expected.

In presence of dots, the mobility is hampered even more, because the potential of the dots causes additional scattering, even though the dots are not charged. The effect is most pronounced for barrier thickness $x=30$\,nm, see Fig.\,\ref{fig:mobility}, where $\mu$ is about an order of magnitude lower in presence of dots. For increasing tunnel barrier widths, the mobility suppression becomes weaker again despite the dots and seems to approach the mobility without dots at the largest barrier thicknesses. We find qualitatively similar results after illumination with a GaAs LED (persistent photoconductivity).

Next, we compare low-$T$ and high-$T$ samples. A higher InAs growth temperature could potentially lead to an enhancement of the 2DEG quality because of two effects: (i) the dot nucleation probability is reduced, resulting in the growth of larger but less dots (lower dot density) \cite{ledentsov:1996, saito:1999}. (ii) The quality of the GaAs capping layer is enhanced (less point defects) due to better annealing. However, comparing upper and lower panels of Fig.\,\ref{fig:mobility}, there is no evidence for a significant improvement. Only the $x=$\,15\,nm point with dots shows slightly better mobility when grown at higher-$T$. We note that with the AFM images, we could not verify a change in size and density of the dots.

These results are in qualitative agreement with previous work on similar inverted hybrid 2DEGs \cite{Sakaki1995}, but the mobility of our material is much higher. Further, we do not observe an increased 2DEG carrier density for small $x$, an undesirable effect seen in a previous study\cite{Sakaki1995}. Thus, the wafers presented here demonstrate a high degree of control resulting in a significant improvement of the hybrid material. Also, a simple WKB estimate of the on-resonance dot-2DEG tunneling rate gives values above 100\,kHz even for the 60\,nm barrier.

\begin{figure}
\centering
\includegraphics[width=1\columnwidth]{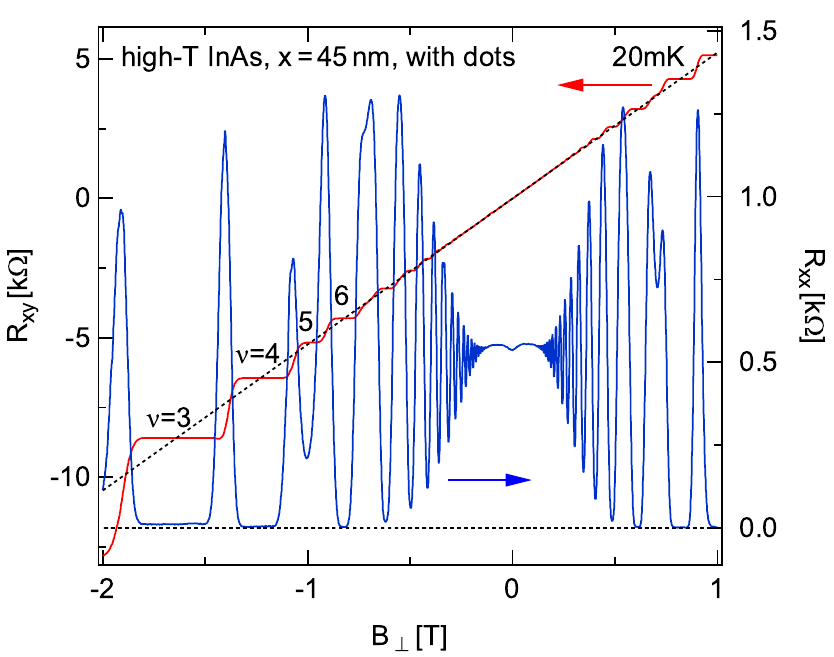}\vspace{-2mm}
\caption{{\bf Shubnikov-de Haas oscillations and quantum Hall effect.} Transverse (red, left axis) and longitudinal (blue, right axis) resistance as a function of perpendicular magnetic field $B_{\bot}$. The density extracted from the classical Hall slope (black dashed line) and from Shubnikov-de Haas oscillations agree (single frequency in $1/B$ before Zeeman splitting appears). Further, the $R_{\rm XX}$-minima go to zero at high $B_{\bot}$, therefore a parallel conduction channel is excluded.\label{fig:SdH}}\vspace{-4mm}
\end{figure}

To further study the 2DEG quality, we perform measurements in a dilution refrigerator at 20\,mK. Fig.\,\ref{fig:SdH} displays longitudinal resistance $R_{\rm XX}$ (blue) and Hall resistance $R_{\rm XY}$ (red) as a function of perpendicular magnetic field $B$. The Fourier transform of the Shubnikov-de-Haas oscillations plotted against $1/B$ results in a single peak for fields below a visible Zeeman splitting, and gives the same 2DEG carrier density as obtained from the classical Hall slope (black dotted line). Further, the Hall resistance is consistent with the extrapolated linear behavior everywhere, and the $R_{\rm XX}$ minima go to zero at the Hall plateaus, which can be fully accounted for with integer filling factors for quantizing fields. All of these signatures testify a high-quality 2DEG despite the presence of dots, without parallel conduction layer and without second subband population.

Around zero field we note a dip in magnetoresistance which we only see in samples with dots. A similar feature has been reported in 2D hole gases\cite{papadakis:2002} and in 2DEGs in presence of a (quasi) periodic potential \cite{weiss:1989, beton:1990, akabori:2000}. Such a potential could be induced by the dots in our samples, possibly giving rise to the zero-field dip.

\begin{figure}
\centering
\includegraphics[width=0.9\columnwidth]{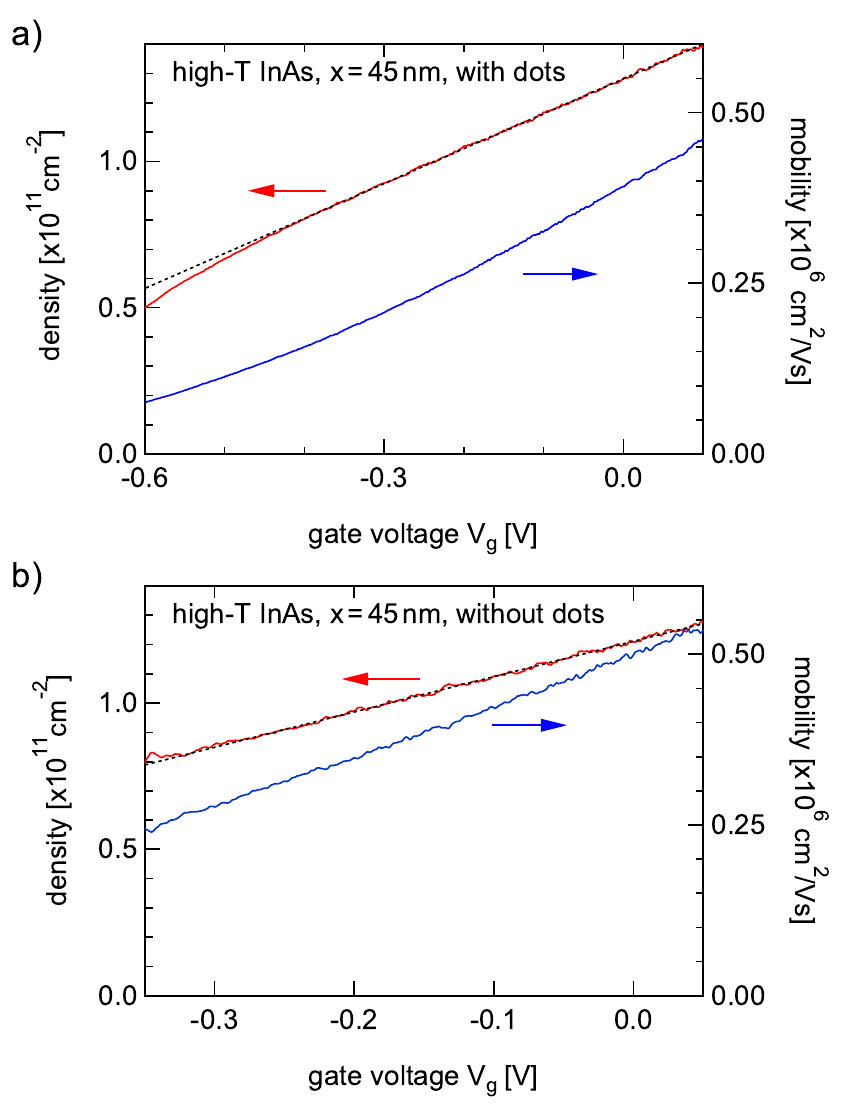}\vspace{-2mm}
\caption{2DEG carrier density $n$ (red, left axis) and mobility $\mu$ (blue, right axis) of 45\,nm tunnel-barrier devices {\bf(a)} in presence and {\bf(b)} in absence of dots, as a function of top gate voltage $V_{\rm g}$. Good gate-tunability is apparent. No significant hysteresis is seen. From linear fits to $n(V_{\rm g})$ (black dashed lines), the 2DEG depth $\approx$500\,nm is extracted, in good agreement with the wafer growth profile.\label{fig:densmob}} \vspace{-4mm}
\end{figure}

Finally, we investigate gateability: a section of the Hall bar is covered with a Ti/Au gate, see Fig.\,\ref{fig:AFM}a). The 2DEG density $n$ and mobility $\mu$ of all devices show an approximately linear gate voltage $V_{\rm g}$ dependence over a relatively wide voltage range, as shown in Fig.\,\ref{fig:densmob} for a high-$T$ wafer with 45\,nm barrier both with dots in a) and without dots in b). In a parallel-plate capacitor model, the capacitance $C$ per area $A$ is given by
\begin{equation}
\frac{C}{A}=\frac{\epsilon_0\epsilon}{d}=e\frac{\partial n}{\partial V_{\rm g}},
\end{equation}
where $d$ is the 2DEG-gate distance, $e>0$ the electron charge, $\epsilon_0$ the vacuum permittivity, and $\epsilon\approx12$ the GaAs dielectric constant. From the slope of a linear fit to $n(V_{\rm g})$, see dashed lines in Fig.\,\ref{fig:densmob}, we obtain $d\approx\,500$\,nm, in good agreement with the as-grown distance $d$. The quantum capacitance of the 2DEG and the capacitance of the measurement setup add in series\cite{Russ2006} and can be neglected, because they are both much larger.

The range of usable top-gate voltages $V_{\rm g}$ is limited by gate-leakage currents and not identical for all devices due to varying mesa etch depths: a smaller etch depth results in a reduced gate voltage range. We choose etch depths in the range of 290-460\,nm, well above the heterointerface and the $\delta$-doping layer. Ohmic contacts experienced adhesion problems when the mesa was etched more deeply into the AlGaAs layer, presumably due to Al oxidation. Dot charging was not accessible in the available gate range.

In summary, we have grown and characterized high quality inverted 2DEG structures with narrow-linewith optical InAs dots in close proximity. AFM scans clearly show the deformation field of the dots -- despite their large distance to the surface. Reduced temperature during InAs growth and presence of InAs dots both reduce the mobility -- more strongly so at small tunnel barrier thicknesses. Nevertheless, high mobilities exceeding $0.5\cdot10^6$cm$^{2}$/(Vs) are obtained while 2DEG-dot tunneling should still be possible. Thus, we have demonstrated crucial ingredients towards future coherent optoelectronics on hybrid systems integrating optically active InAs dots with accurately placed nanostructures in a high-quality 2DEG.

\begin{acknowledgments}
We thank A. Imamoglu for valuable discussions. This work was supported by the Swiss Nanoscience Institute (SNI), NCCR QSIT, Swiss NSF, ERC starting grant (D.M.Z.), EU-FP7 SOLID, and ERA Project QOptInt (E.R.).
\end{acknowledgments}


%

\end{document}